\documentclass[sigconf,nonacm]{acmart}
\AtBeginDocument{%
  }

\setcopyright{acmlicensed}
\copyrightyear{2026}
\acmYear{2026}
\acmDOI{X}
\acmConference[CHI '26]{CHI Conference on Human Factors in Computing Systems Workshop on Data Literacy}{April 13-17, 2026}{Barcelona}
\acmISBN{978-1-4503-XXXX-X/2026/05}

\newcommand{\edit}[1]{#1}

\usepackage{pifont}
\usepackage{multirow}
\usepackage[dvipsnames]{xcolor} 

\newcommand{\gtrain}{\textit{Learn Like an LLM}}
\newcommand{\gpredict}{\textit{Tag-Team Text Generation}}

\newcommand{\figref}[1]{Fig.~\ref{#1}}

\newcommand{\appendixref}[1]{Appendix}
\newcommand{\appendixrefA}[1]{Appendix Sec. A}
\newcommand{\appendixrefB}[1]{Appendix Sec. B}
\newcommand{\appendixrefC}[1]{Appendix Sec. C}

\begin{document}

\title{Using Games to Learn How Large Language Models Work}

\author{Allison Chen}
\email{allisonchen@princeton.edu}
\orcid{0000-0002-0919-1849}
\affiliation{%
  \institution{Princeton University}
  \city{Princeton}
  \state{New Jersey}
  \country{USA}
}

\author{Isabella Pu}
\email{ipu@media.mit.edu}
\orcid{0009-0009-0798-0949}
\affiliation{%
  \institution{Massachusetts Institute of Technology}
  \city{Cambridge}
  \state{Massachusetts}
  \country{USA}
}


\renewcommand{\shortauthors}{Chen \& Pu}

\begin{abstract} 
    While artificial intelligence (AI) technology is becoming increasingly popular, its underlying mechanisms tend to remain opaque to most people.
    To address this gap, the field of AI literacy aims to develop various resources to teach people how AI systems function.
    Here we contribute to this line of work by proposing two games that demonstrate principles behind how large language models (LLMs) work and use data.
    The first game, \gtrain, aims to convey that LLMs are trained to predict sequences of text based on a particular dataset.
    The second game, \gpredict, focuses on teaching that LLMs generate text one word at a time, using both predicted probabilities of the data and randomness.
    While the games proposed are still in early stages and would benefit greatly from further discussion, we hope they can contribute to using game-based learning to teach about complex AI systems like LLMs. \looseness=-1
\end{abstract}




\received{10 February 2026}

\maketitle

\section{Introduction}

As artificial intelligence (AI) technology becomes increasingly prevalent, there is a need for people to develop an accurate understanding about these systems, including how they work and how to use them.
Since ChatGPT released in 2022, AI has had a wide reach, affecting economics \citep{wang2025does, farina2025ethical}, politics and policy \citep{best2024future}, and daily life \citep{trujillo2025global, chatterji2025people, mcdaniel2025artificial, brandtzaeg2025emerging}.
However, despite its growing prevalence, many people have limited access to the resources and training needed to develop an accurate understanding about the AI technology.
This gap is further widened by the fact that many AI systems produce human-sounding text (e.g., using first-person pronouns) and many organizations describe the systems as if they were human, using phrases like AI can ``learn'' or ``possesses intelligence'' \citep{devrio2025taxonomy, salles2020anthropomorphism}. In other words, there is a lot of anthropomorphic language around AI systems.
While this use of anthropomorphic words may make complex AI technology \textit{feel} more approachable, it simultaneously masks the technical and algorithmic nature of these systems. As a result, this can lead to increased risks of people developing emotional attachments and overestimating the system's capabilities \citep{alasgarova2025managing, cheng2024one, li2025surprising}. \looseness=-1

To help people develop a better understanding about AI technology, the field of AI literacy seeks to develop educational materials and resources that combat misleading anthropomorphic narratives about AI \citep{ng2021conceptualizing, long2020ai, lintner2024systematic}. 
AI literacy encompasses many aspects of how people interact with and relate to AI systems, such as what it can do, how it works, how it should be used, and how people perceive these systems \citep{long2020ai}. 
\edit{While AI literacy draws from many other forms of literacy, it is closely connected to data literacy.
For example, data literacy primarily focuses on how to collect, manage, evaluate, and apply data critically \citep{ridsdale2015strategies}, and a large part of AI literacy can draw upon these ideas to demonstrate how AI systems leverage large amounts of data. The process by which AI systems pick up on statistical patterns is called ``training.''
Understanding \textit{how} this data can shape an AI system's capabilities and predictions is a critical component of AI literacy.} 
This is important in order to equip people with the tools to make informed decisions, such as financial investments, regulation and policy, and everyday use at work or home \citep{fard2026growing}.
Further, it can help reduce both the sense of ``mystery'' surrounding AI technology and people's overreliance on these systems \citep{wood2025exploratory, alasgarova2025managing, tully2025lower}
\looseness=-1

In this work, we build upon existing works promoting AI literacy through informal learning settings by outlining novel ideas for two games.
These games aim to intuitively convey fundamental ideas behind how modern AI systems \edit{work. Specifically, we focus on how large language models (LLMs) are trained and generate text.}
We chose to prioritize LLMs because they are a popular AI system and many people use them, either directly through chat interfaces or indirectly through other software products (e.g., Google search) \citep{liang2025quantifying, brachman2025current}.
Prior AI literacy works tended to focus on foundational concepts like machine learning or neural networks \citep{carney2020teachable, smilkov2017direct}, but it is becoming increasingly important to also develop materials about specific AI systems that people may encounter frequently, such as LLMs \citep{ma2025imaginaition, annapureddy2025generative}. 
While there are a few works that focus on generative AI systems like LLMs, these primarily aim to explain the behaviors or impact of the systems, not necessarily how they work \edit{and use data} \citep{cao2025empowering, ma2025imaginaition, laupichler2022artificial}. 
To help fill this gap, our proposal is centered around two games designed to convey technical concepts about the mechanisms of how LLMs produce text.

While most avenues of AI and data literacy occur in formal classroom settings \citep{casal2023ai, su2023artificial, laupichler2022artificial, cui2023data, gebre2022conceptions}, our work contributes to the growing literature that recognizes the value of developing literacy activities and resources for informal settings, specifically through games.
Learning in informal environments, such as museums or other public spaces \citep{long2021co, darabipourshiraz2026ai, darabipourshiraz2024databites, rollins2024knowledge, sager2026data} recognizes these low-stakes settings tend to be more engaging, allow parents and children to learn together, and make information more accessible.
Games specifically have been shown to be effective learning tools \citep{plass2015foundations, tobias2013game} for adults as well as children \citep{charlier2012not}.
First, games can be motivating, engaging, and adaptive to different learning styles \citep{plass2015foundations}.
They further provide low-stakes opportunities for experimentation and feedback, creating a safe space for failure that is important to effective learning \citep{ chow_deal_2011, zeng_learn_2020}.
Games also help foster autonomy and curiosity, making abstract or intimidating concepts like AI more approachable through playful exploration and immediate feedback \citep{nand_engaging_2019, rondon_computer_2013, giannakos_games_2020}.
\edit{In both AI and data} literacy, games have been increasingly employed and shown to effectively teach AI-related concepts \citep{du2024fostering, ng2024fostering, darabipourshiraz2026ai, rollins2024knowledge} while additionally providing positive affective experiences \citep{ng2024fostering, long2022family}.
Here, we build upon game-based learning to develop AI literacy materials regarding specifically how LLMs work and use data.

We propose two games with the goal of filling the gap in the public's familiarity with and understanding of LLM systems.
The first game focuses on \edit{a core data literacy concept:} how LLMs learn to generate text via predicting sequences of words from a \edit{particular} dataset.
The second focuses on how LLMs generate text one word at a time, influenced by \edit{data, prediction probabilities, and randomness.} \looseness=-1

In first game, \gtrain, players are tasked with guessing sequences of shapes that belong to a hidden set, similar to how LLMs are tasked to predict sequences of text from the training dataset.
The primary learning objectives is for players to understand \edit{how the LLM training process is analogous to pattern recognition with a very large dataset, and that the content and composition of that dataset mediate the model's behavior. Secondarily, we hope players understand that the system is sensitive to the quality and content of the data.} \looseness=-1

In the second game, \gpredict, players take turns with a ``computer player'' to generate responses one word at a time to fun open-ended questions.
Here, the primary learning objective is for players to recognize that LLMs do not ``think about ideas'' then find the words to convey them like people do. 
Rather, LLMs generate text by predicting one word\footnote{Truly, it is a sub-word token, but for the purposes of our game, we do not go into this detail.} at a time.
\edit{We additionally aim to convey that LLMs predict text by indirectly leveraging probabilities learned from the training data but also incorporate randomness which reduces the influence of training data patterns on its final predictions.}
We note that these games do not aim to provide people with a \textit{complete} technical understanding of LLMs.
Rather, they aim to give players an \textit{intuition} about specific aspects of LLMs that lead them to say \textit{``This isn't how I thought these technologies worked!''} or \textit{``This is not so complicated, I can understand this!''} 

\section{Game 1: \gtrain}

The main goal of \gtrain \ is to help people understand that LLMs are optimized to predict words that are likely to come next given a sequence of text and it is heavily dependent on its particular training dataset.
To convey this idea, players are given a ``vocabulary'' of 10-12 simple shapes of different colors, and their goal is to guess which sequences of shapes are in the hidden ``training'' set\footnote{The mechanics of the game are similar to Mastermind or Wordle.}.
Then they are given the following instructions:

\begin{figure*}[t]
    \centering
    \fbox{
    \includegraphics[width=10.075cm,trim={0 0 0 0},clip]{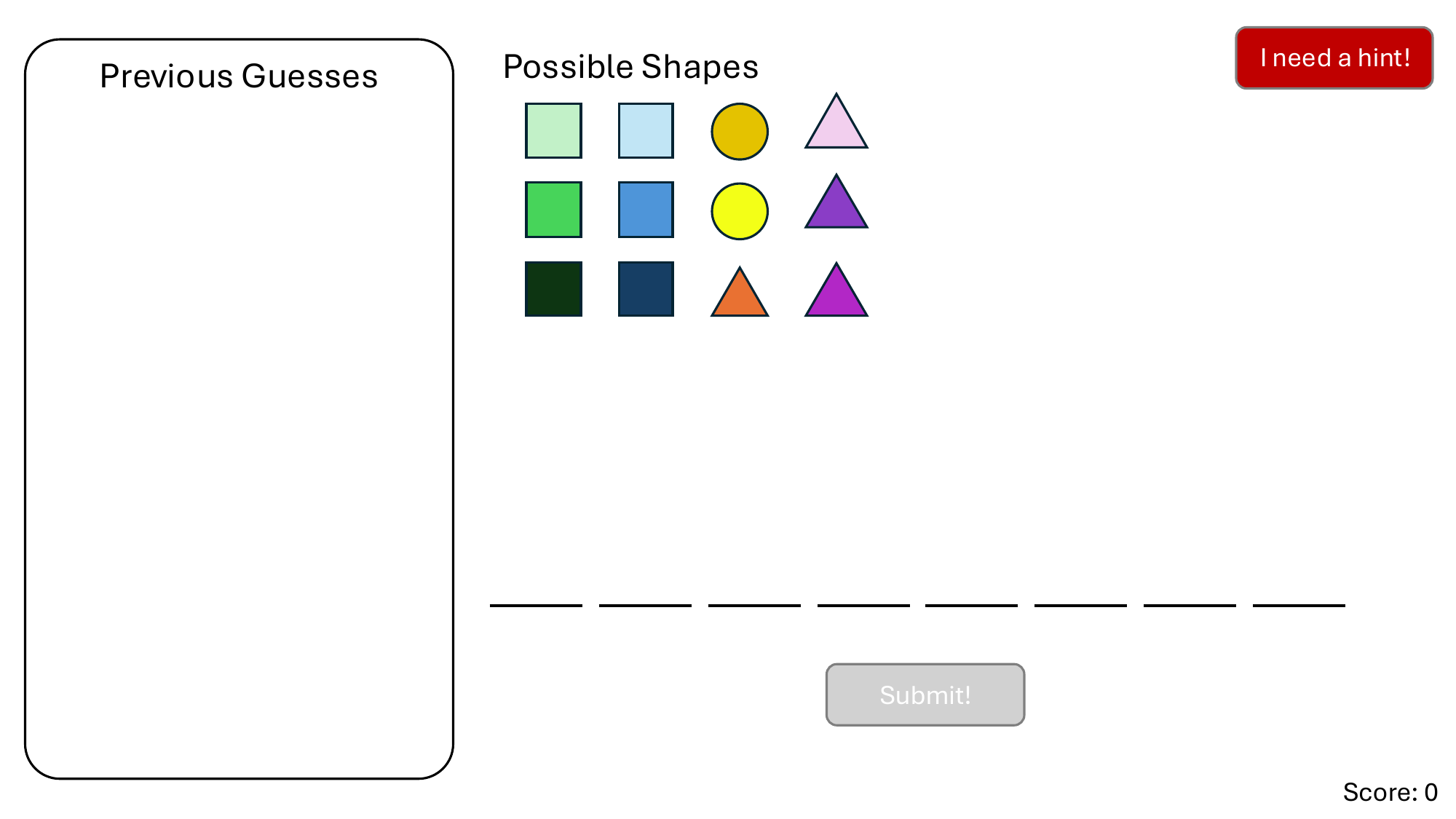}}
    \fbox{
    \includegraphics[width=10cm,trim={0 0 0 0},clip]{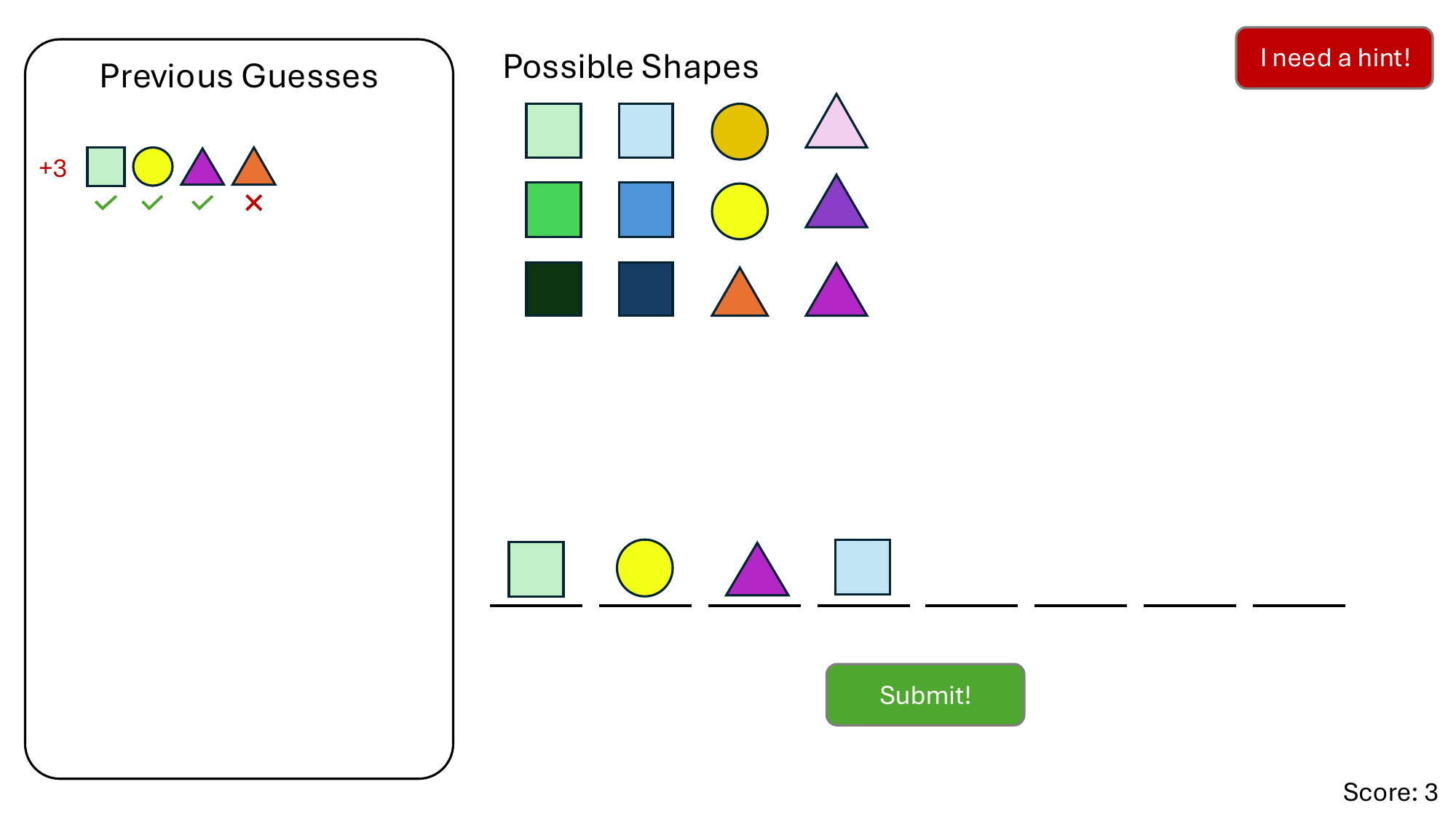}
    }
        \caption{Two example interfaces of Game 1 (\gtrain). \textbf{Top}: This is the interface when the player starts. The set of possible shapes is always at the top of the screen. Once the player selects at least 4 shapes, the Submit button becomes clickable. \textbf{Bottom}: A possible interface after the player submits one sequence and is constructing the second. The submitted sequene is on the left side of the screen. The points earned from the first sequence is to the left of the sequence, the color of the points denotes whether the shape was in the hidden set, and the validity of each shape is denoted by the checkmarks and X's under each shape.}
        
\label{fig:game1_interface}
\end{figure*}

\begin{enumerate}
    \item There is a hidden set of sequences of shapes. Each sequence is between 4-8 shapes long, and they follow some patterns or rules. Your goal is to guess as many sequences in the hidden set as possible!
    \item You can build a sequence by selecting one shape at a time, and you may submit your guess for a sequence once you have selected at least 4 shapes. Or you may continue to build the sequence up to 8 shapes long. 
    \item When the sequence is submitted, you may receive points for each shape in the sequence and for the whole sequence. You will receive one point for each shape that is valid given the preceding shape (or if it is the first shape, whether it is a valid first shape). Second, if all shapes are valid \textit{and} the sequence was in the hidden set, you will receive additional 3 points for guessing the sequence. Note, not all valid sequences are in the hidden set, \textit{but} this can still be helpful information to help you deduce the rules governing the sequences.
    \item If you get stuck, you may click on the ``I need a hint!'' button, and this will reveal a sequence from the hidden set. However, this counts as one of your guesses and you will not receive any points from it.
    \item You have 10 tries to guess as many sequences as you can!
\end{enumerate}

\begin{figure*}[t]
    \centering
    \fbox{
    \includegraphics[width=0.7\linewidth,trim={0 5.5cm 0 4.5cm},clip]{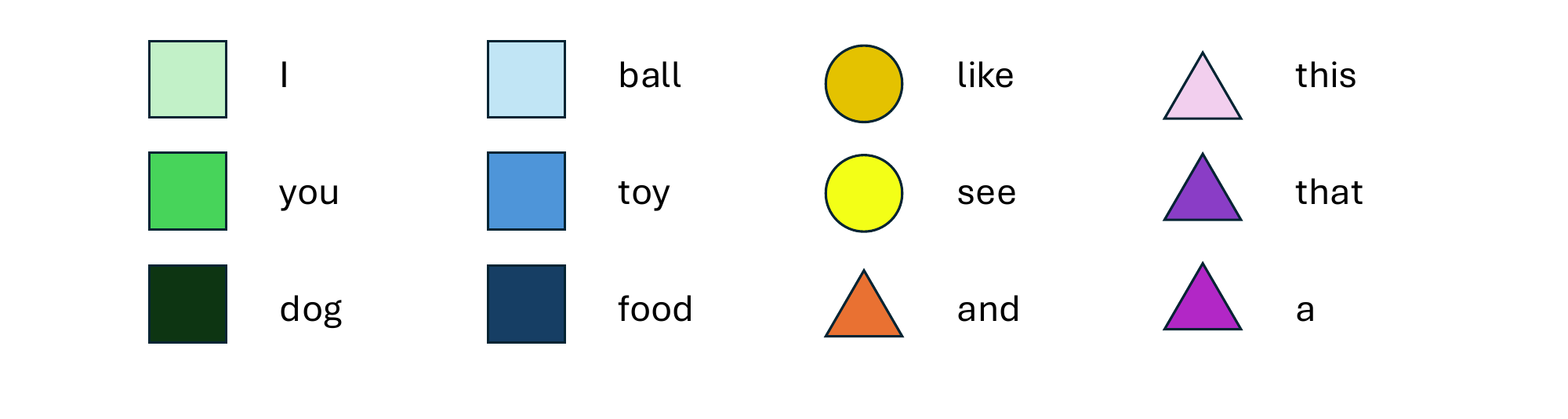}}
    \caption{Possible mapping of shapes to words for \gtrain. Based on this mapping, the previously submitted sequence in \figref{fig:game1_interface} (bottom) maps to ``I see a and'' and the sequence in progress maps to ``I see a ball''.}
\label{fig:game1_mapping}
\end{figure*}

After reading the instructions, the player will see the game interface shown in \figref{fig:game1_interface} (top) where they can guess which sequences of shapes belong to the hidden set.
The current sequence will be built on the dashed lines and the ``Submit!'' button is valid once there is a minimum of four shapes selected.
When the player submits a sequence, it will appear on the left hand side with the points received, as shown in \figref{fig:game1_interface} (bottom). 
The ``\textcolor{red}{+3}'' denotes the player earned 3 points from that sequence, one for each shape with a green checkmark under it indicating that it was valid. The red color of the ``\textcolor{red}{+3}'' indicates that the sequence was not in the hidden set. 
The game ends after the player submits 10 guesses. 
Unbeknownst to the player, each shape actually represents an English word and the sequences are based on grammatically correct sentences\footnote{Barring conjugations for tense and plurality.}. An example shape-to-word mapping can be found in \figref{fig:game1_mapping}.
Because the sequences mirror English grammar, the goal is that players will pick up on patterns, such as how certain shapes tend to follow others. \looseness=-1

After the game ends, this shape-to-word mapping will be revealed and the player will receive a debriefing on how the game mirrors the way LLMs are trained, \edit{model patterns in the training data, and can be susceptible to reinforcing biases in the dataset.}
Primarily, the hidden set reflects training data that LLMs learn to predict the next word given a sequence of text. 
Similarly, in this game, the players have a sequence of shapes that they have to predict one shape at a time.
Further, at the start of the game, the shapes had no semantic meaning or relation to one another for the player.
However, over the course of receiving feedback for correct and incorrect guesses, the player may learn associations between certain shapes without necessarily ascribing meaning to the shapes. 
This mirrors how before training, words do not carry semantic meanings to LLMs, but as the LLM is exposed to more combinations of words and sentences in the data, it can implicitly model statistical patterns between. 
We hope this point raises the question for the player to ponder and discuss: even if LLMs can robustly model the statistical relations of words, does this mean it knows the words' meaning?
Lastly, the hidden set of sequences did not contain \textit{all} grammatically correct combinations of the shapes, which meant that the player is penalized for valid English sequences beyond those in the set. 
This is similar to how LLMs have non-exhaustive training datasets and can be constrained by their training data. 
Practically, this can lead LLMs to pick up biases in the training data. 
In summary, \gtrain \ aims to players how LLMs are reliant on their data. 
LLMs are trained to generate text by predicting sequences of text from a particular dataset and are penalized for wrong predictions and rewarded for correct predictions.

\section{Game 2: \gpredict}

The second game, \gpredict, is a creative and non-competitive activity that aims to convey how LLMs generate text one word at a time\footnote{As mentioned in the introduction, in practice, LLMs generate text one \textit{token} at a time where tokens can be sub-words or full words, but we omit this detail in the game for brevity.} and how each word prediction utilizes both probabilities \edit{implicitly learned from the dataset} and randomness.
In this game, the player will alternate with the computer to simulate how LLMs generate responses: one word at a time and through a two-step process using probabilities and randomness. 

The player starts by selecting a fun prompt to answer, such as one of the following:

\begin{itemize}
    \item What would Cinderella’s godmother give her if she lived in 2026?
    \item Describe a sunset on the beach from the perspective of a baby turtle who just hatched.
    \item Create your own fantasy creature, describing what it looks like and what it likes to do.
    \item Where is the best place to live---real or imaginary---and why?
    \item If you were a ruler of a country, what would be your national food and why?
\end{itemize}

\begin{figure*}[t]
    \centering
    \fbox{
    \includegraphics[width=0.7\linewidth,trim={0 0 0 0cm},clip]{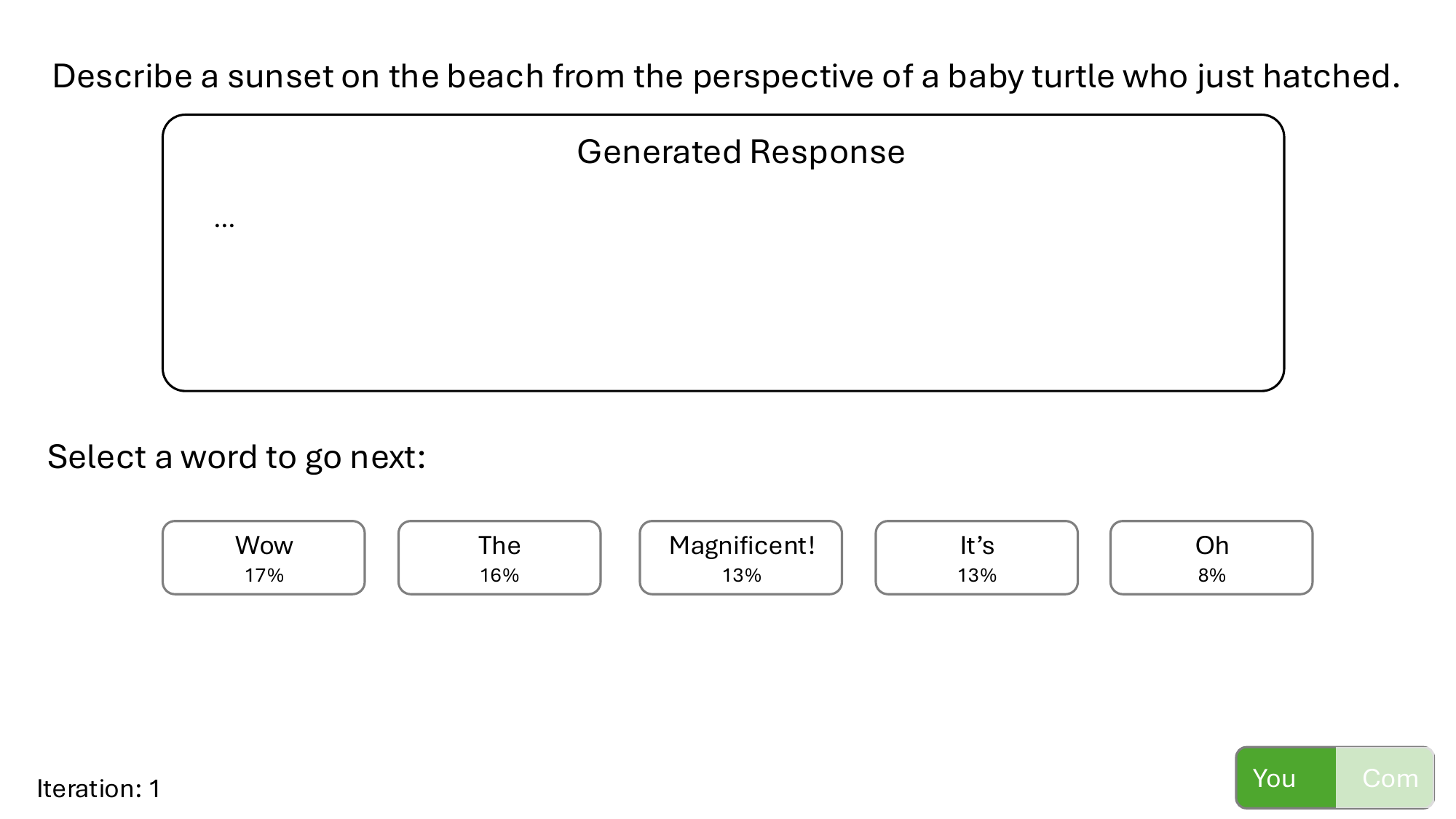}}
    \fbox{
    \includegraphics[width=0.7\linewidth,trim={0 0 0 0cm},clip]{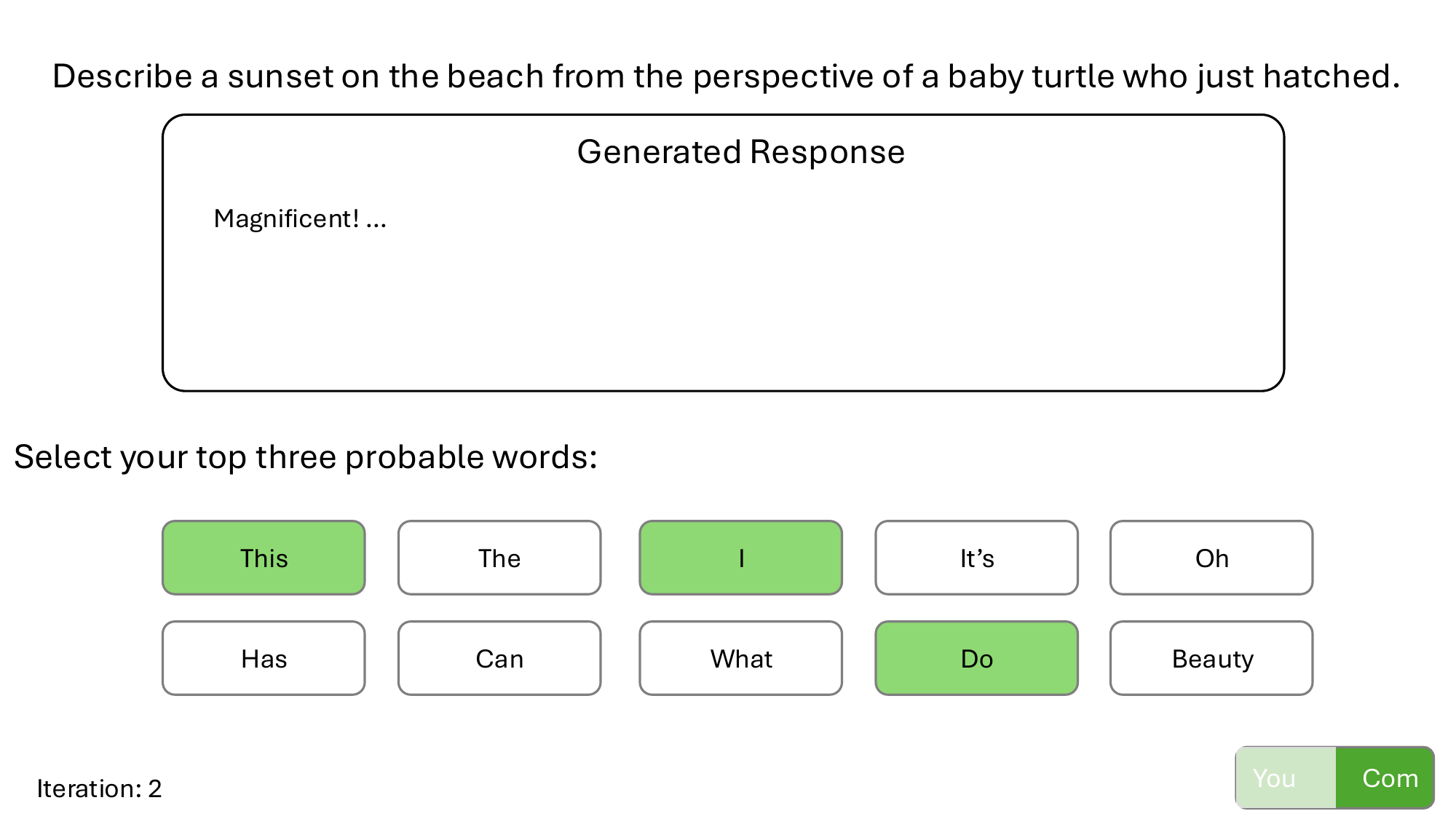}}
    \caption{Interface for \gpredict. \textbf{Top}: Interface when it is the player's turn to make the final word selection from a set of 5 words with their estimated probabilities. Here they select one. \textbf{Bottom}: Interface when it is the player's turn to generate a set of probable words. In this example, the player opted to select three from a pool of 10, rather than submitting their own. The computer will select one of the three selected words to add to the response.}
        
\label{fig:game2}
\end{figure*}

Alternatively, players can also submit their own prompt. Then to build the response, the player and computer will alternate between generating a set of probable words and selecting a word from the generated set.

To start, the computer will generate and display a set of 5 possible words with their corresponding probabilities, such as shown in \figref{fig:game2} (top). 
The probabilities do not necessarily need to add up to 100\% as these words only represent a \textit{subset} of all possible options.
The player then selects any word they like---it could be random, based on what they think is funny, etc. 
They do not need to choose the one that is most probable, and in fact, varying their selections will likely be more entertaining.
Once the player selects the word, it is added to the response as shown in \figref{fig:game2} for the word ``Magnificent!'' \looseness=-1

To generate the next word, the roles are flipped. Now, the player will provide the computer with a small set of possible words that the computer will randomly select from.
To do so, the player can either choose to come up with three words they think are most probable \textit{or} select three from a larger pool of possible words if they feel this is too cumbersome. 
If they select the latter option, the player is presented with 10 possible words (without any probabilities), as shown in \figref{fig:game2} (bottom). 
The computer will then take the three submitted words and randomly select one of them as the next word in the response. 
Because the computer's selection process is agnostic to the words provided, if an adversarial player tries to submit non-sensical words, the computer will still select one, but the subsequent generated words from the computer will likely degrade.
\looseness=-1

These two steps---where the player selects a final word or creates a pool of probable words---repeats until the player is satisfied with their response and can submit it. 
Upon submitting, the player can see other people's responses to the same prompt; this will be especially conducive for an online setting as well as in museums and public spaces.
As an additional reward, we can add the option for players to submit their prompt and generated response to a text-to-image generation model and print out the resulting image.

After the player is done, they will receive a debriefing of how this game reflects the text generation mechanisms of LLMs \edit{and how training data influences word-level predictions}. 
First, players experienced needing to generate only one word at a time; this limits how much they could ``plan ahead'' on what to write because they were limited by the computer's estimates of which words were most probable or by the computer's random selection. 
The players may have experienced hoping to take the response in one direction but the computer's selection was incompatible with their plan.
Similarly, LLMs do not ``think ahead'' about what text to generate.
Second, players experienced a two-step process: selecting a pool of possible words based on probabilities and choosing a word from that pool.
LLMs undergo a similar process when they generate words.
\edit{They first compute how probable each word is which is largely based on the patterns learned from the training data.}
Then to generate text that can be more diverse than those in the training data, they use an algorithm that leverages some randomness to select the final prediction for each word.
This randomness is what allows for diverse responses to the same prompt.
Conveying that randomness is a part of the text generation process further emphasizes the idea from \gtrain \ that LLMs do not generate text by coming up with ideas to then convey through text.
In summary, \gpredict \ aims to demonstrate to players that LLMs predict text one word at a time through a process that involved both probabilities \edit{implicitly learned from the training data} and randomness.

\section{Discussion}

\subsection{Summary}

In this work, we propose two games, \gtrain \ and \gpredict \ that aim to convey how LLMs work, \edit{specifically how data is used during training and how LLMs generate text.} 
In the first game, \gtrain, players are taught that LLMs are trained to produce text like how  one can predict sequences of shapes using pattern recognition. 
The game aims to reinforce that LLMs do not learn to generate text the same way people learn to use words to communicate ideas, but rather that LLMs operate based on pattern recognition and modeling. 
\edit{Further, it engages players in core data literacy reasoning by asking them to examine the composition of a dataset and recognize how biases or gaps in the data translate to the model's outputs.}

In the second game, \gpredict, players learn that the LLM text generation process can be broken down into two steps: using probabilities to select a subset of words that are likely to come next and selecting one of them with a procedure that uses randomness. 
\edit{This randomness allows LLMs to deviate from sentences and patterns present in the training data and produce seemingly creative sentences.}
\gpredict is meant to reveal that the text-generation procedure is not as sophisticated as many companies may claim, but is based on probabilities and randomness.

\subsection{Limitations}

While these games aim to shed light on some aspects of how LLMs work, we acknowledge a few limitations; primarily, the games still abstract away details of LLMs, they do not address aspects such as how data is represented to LLMs, and may be possible for people to play them without understanding how they relate to LLMs.
First, the games still abstract away certain details and are not perfect reflections of how LLMs are trained and predict text. 
For example, LLMs are provided a sequence of text as an input and are tasked to generate one word, but in \gtrain, players are building up sequences from scratch.
With \gpredict, it seems that LLMs output only a few words that are probable, but in reality, they output a probability distribution over the entire vocabulary.
Future work can iterate on how to incorporate this aspect into one of the proposed games or into an entirely new activity.
A second limitation is that there are many aspects of LLMs that these games do not touch on. For example, they do not shed light on how words are represented as vectors, how text gets split up into units called tokens, or how, in practice, many LLMs are likely to use more sophisticated sampling techniques to select the final word.
We hope that future work can focus on developing games and activities to address these topics.
Lastly, it is possible for people to play these games without ever connecting them to LLMs. 
While this may still provide a fun and engaging experience, we hope to improve the games by highlighting how they are  connected to LLMs before the debriefing at the end.

\subsection{Conclusion}

In this work, we outline early stage ideas for two games conveying how LLMs \edit{leverage data to generate text}: \gtrain focuses on the training procedure and \gpredict on LLM text generation \edit{via word prediction}. 
We hope that the CHI 2026 Workshop on Data Literacy will be a helpful opportunity to receive feedback and actively iterate on these ideas to address the aforementioned limitations, better convey intuitions about how LLMs work, and improve the gameplay experience overall.
We additionally hope that this work will open up conversations for novel and engaging ways of using games to teach people---children and adults alike---about how LLMs and other modern AI systems work through engaging and fun activities!

\subsection*{Acknowledgments}

We thank Vivian Zeng for her help developing ideas and reading over the manuscript and acknowledge support from the NSF GRFP (A.C.).

\bibliographystyle{ACM-Reference-Format}
\bibliography{references}


\begin{thebibliography}{47}


\ifx \showCODEN    \undefined \def \showCODEN     #1{\unskip}     \fi
\ifx \showISBNx    \undefined \def \showISBNx     #1{\unskip}     \fi
\ifx \showISBNxiii \undefined \def \showISBNxiii  #1{\unskip}     \fi
\ifx \showISSN     \undefined \def \showISSN      #1{\unskip}     \fi
\ifx \showLCCN     \undefined \def \showLCCN      #1{\unskip}     \fi
\ifx \shownote     \undefined \def \shownote      #1{#1}          \fi
\ifx \showarticletitle \undefined \def \showarticletitle #1{#1}   \fi
\ifx \showURL      \undefined \def \showURL       {\relax}        \fi
\providecommand\bibfield[2]{#2}
\providecommand\bibinfo[2]{#2}
\providecommand\natexlab[1]{#1}
\providecommand\showeprint[2][]{arXiv:#2}

\bibitem[Alasgarova and Rzayev(2025)]%
        {alasgarova2025managing}
\bibfield{author}{\bibinfo{person}{Rena Alasgarova} {and} \bibinfo{person}{Jeyhun Rzayev}.} \bibinfo{year}{2025}\natexlab{}.
\newblock \showarticletitle{Managing Anthropomorphism in Student--AI Interaction}.
\newblock \bibinfo{journal}{\emph{International Journal of Information Technology and Education}} \bibinfo{volume}{5}, \bibinfo{number}{1} (\bibinfo{year}{2025}), \bibinfo{pages}{139--160}.
\newblock


\bibitem[Annapureddy et~al\mbox{.}(2025)]%
        {annapureddy2025generative}
\bibfield{author}{\bibinfo{person}{Ravinithesh Annapureddy}, \bibinfo{person}{Alessandro Fornaroli}, {and} \bibinfo{person}{Daniel Gatica-Perez}.} \bibinfo{year}{2025}\natexlab{}.
\newblock \showarticletitle{Generative AI literacy: Twelve defining competencies}.
\newblock \bibinfo{journal}{\emph{Digital Government: Research and Practice}} \bibinfo{volume}{6}, \bibinfo{number}{1} (\bibinfo{year}{2025}), \bibinfo{pages}{1--21}.
\newblock


\bibitem[Best et~al\mbox{.}(2024)]%
        {best2024future}
\bibfield{author}{\bibinfo{person}{Eric Best}, \bibinfo{person}{Pedro Robles}, {and} \bibinfo{person}{Daniel~J Mallinson}.} \bibinfo{year}{2024}\natexlab{}.
\newblock \showarticletitle{The Future of AI Politics, Policy, and Business}.
\newblock \bibinfo{journal}{\emph{Business and Politics}} \bibinfo{volume}{26}, \bibinfo{number}{2} (\bibinfo{year}{2024}), \bibinfo{pages}{171--179}.
\newblock


\bibitem[Brachman et~al\mbox{.}(2025)]%
        {brachman2025current}
\bibfield{author}{\bibinfo{person}{Michelle Brachman}, \bibinfo{person}{Amina El-Ashry}, \bibinfo{person}{Casey Dugan}, {and} \bibinfo{person}{Werner Geyer}.} \bibinfo{year}{2025}\natexlab{}.
\newblock \showarticletitle{Current and future use of large language models for knowledge work}.
\newblock \bibinfo{journal}{\emph{Proceedings of the ACM on Human-Computer Interaction}} \bibinfo{volume}{9}, \bibinfo{number}{7} (\bibinfo{year}{2025}), \bibinfo{pages}{1--24}.
\newblock


\bibitem[Brandtzaeg et~al\mbox{.}(2025)]%
        {brandtzaeg2025emerging}
\bibfield{author}{\bibinfo{person}{Petter~Bae Brandtzaeg}, \bibinfo{person}{Asbj{\o}rn F{\o}lstad}, {and} \bibinfo{person}{Marita Skjuve}.} \bibinfo{year}{2025}\natexlab{}.
\newblock \showarticletitle{Emerging AI Individualism: How Young People Integrate Social AI into Everyday Life}.
\newblock \bibinfo{journal}{\emph{Communication and Change}} \bibinfo{volume}{1}, \bibinfo{number}{1} (\bibinfo{year}{2025}), \bibinfo{pages}{11}.
\newblock


\bibitem[Cao et~al\mbox{.}(2025)]%
        {cao2025empowering}
\bibfield{author}{\bibinfo{person}{Huajie~Jay Cao}, \bibinfo{person}{Hee~Rin Lee}, {and} \bibinfo{person}{Wei Peng}.} \bibinfo{year}{2025}\natexlab{}.
\newblock \showarticletitle{Empowering Adults with AI Literacy: Using Short Videos to Transform Understanding and Harness Fear for Critical Thinking}. In \bibinfo{booktitle}{\emph{Proceedings of the 2025 CHI Conference on Human Factors in Computing Systems}}. \bibinfo{pages}{1--8}.
\newblock


\bibitem[Carney et~al\mbox{.}(2020)]%
        {carney2020teachable}
\bibfield{author}{\bibinfo{person}{Michelle Carney}, \bibinfo{person}{Barron Webster}, \bibinfo{person}{Irene Alvarado}, \bibinfo{person}{Kyle Phillips}, \bibinfo{person}{Noura Howell}, \bibinfo{person}{Jordan Griffith}, \bibinfo{person}{Jonas Jongejan}, \bibinfo{person}{Amit Pitaru}, {and} \bibinfo{person}{Alexander Chen}.} \bibinfo{year}{2020}\natexlab{}.
\newblock \showarticletitle{Teachable machine: Approachable Web-based tool for exploring machine learning classification}. In \bibinfo{booktitle}{\emph{Extended abstracts of the 2020 CHI conference on human factors in computing systems}}. \bibinfo{pages}{1--8}.
\newblock


\bibitem[Casal-Otero et~al\mbox{.}(2023)]%
        {casal2023ai}
\bibfield{author}{\bibinfo{person}{Lorena Casal-Otero}, \bibinfo{person}{Alejandro Catala}, \bibinfo{person}{Carmen Fern{\'a}ndez-Morante}, \bibinfo{person}{Maria Taboada}, \bibinfo{person}{Beatriz Cebreiro}, {and} \bibinfo{person}{Sen{\'e}n Barro}.} \bibinfo{year}{2023}\natexlab{}.
\newblock \showarticletitle{AI literacy in K-12: a systematic literature review}.
\newblock \bibinfo{journal}{\emph{International Journal of STEM Education}} \bibinfo{volume}{10}, \bibinfo{number}{1} (\bibinfo{year}{2023}), \bibinfo{pages}{29}.
\newblock


\bibitem[Charlier et~al\mbox{.}(2012)]%
        {charlier2012not}
\bibfield{author}{\bibinfo{person}{Nathalie Charlier}, \bibinfo{person}{Michela Ott}, \bibinfo{person}{Bernd Remmele}, {and} \bibinfo{person}{Nicola Whitton}.} \bibinfo{year}{2012}\natexlab{}.
\newblock \showarticletitle{Not just for children: game-based learning for older adults}. In \bibinfo{booktitle}{\emph{6th European Conference on Games Based Learning, Cork, Ireland}}. \bibinfo{pages}{102--108}.
\newblock


\bibitem[Chatterji et~al\mbox{.}(2025)]%
        {chatterji2025people}
\bibfield{author}{\bibinfo{person}{Aaron Chatterji}, \bibinfo{person}{Thomas Cunningham}, \bibinfo{person}{David~J Deming}, \bibinfo{person}{Zoe Hitzig}, \bibinfo{person}{Christopher Ong}, \bibinfo{person}{Carl~Yan Shan}, {and} \bibinfo{person}{Kevin Wadman}.} \bibinfo{year}{2025}\natexlab{}.
\newblock \bibinfo{booktitle}{\emph{How People Use ChatGPT}}.
\newblock \bibinfo{type}{{T}echnical {R}eport}. \bibinfo{institution}{National Bureau of Economic Research}.
\newblock


\bibitem[Cheng et~al\mbox{.}(2024)]%
        {cheng2024one}
\bibfield{author}{\bibinfo{person}{Myra Cheng}, \bibinfo{person}{Alicia DeVrio}, \bibinfo{person}{Lisa Egede}, \bibinfo{person}{Su~Lin Blodgett}, {and} \bibinfo{person}{Alexandra Olteanu}.} \bibinfo{year}{2024}\natexlab{}.
\newblock \showarticletitle{" I Am the One and Only, Your Cyber BFF": Understanding the Impact of GenAI Requires Understanding the Impact of Anthropomorphic AI}.
\newblock \bibinfo{journal}{\emph{arXiv preprint arXiv:2410.08526}} (\bibinfo{year}{2024}).
\newblock


\bibitem[Chow et~al\mbox{.}(2011)]%
        {chow_deal_2011}
\bibfield{author}{\bibinfo{person}{Alan~F. Chow}, \bibinfo{person}{Kelly~C. Woodford}, {and} \bibinfo{person}{Jeanne Maes}.} \bibinfo{year}{2011}\natexlab{}.
\newblock \showarticletitle{Deal or {No} {Deal}: using games to improve student learning, retention and decision-making}.
\newblock \bibinfo{journal}{\emph{International Journal of Mathematical Education in Science and Technology}} \bibinfo{volume}{42}, \bibinfo{number}{2} (\bibinfo{date}{March} \bibinfo{year}{2011}), \bibinfo{pages}{259--264}.
\newblock
\showISSN{0020-739X, 1464-5211}
\href{https://doi.org/10.1080/0020739X.2010.519796}{doi:\nolinkurl{10.1080/0020739X.2010.519796}}


\bibitem[Cui et~al\mbox{.}(2023)]%
        {cui2023data}
\bibfield{author}{\bibinfo{person}{Ying Cui}, \bibinfo{person}{Fu Chen}, \bibinfo{person}{Alina Lutsyk}, \bibinfo{person}{Jacqueline~P Leighton}, {and} \bibinfo{person}{Maria Cutumisu}.} \bibinfo{year}{2023}\natexlab{}.
\newblock \showarticletitle{Data literacy assessments: a systematic literature review}.
\newblock \bibinfo{journal}{\emph{Assessment in education: Principles, policy \& Practice}} \bibinfo{volume}{30}, \bibinfo{number}{1} (\bibinfo{year}{2023}), \bibinfo{pages}{76--96}.
\newblock


\bibitem[Darabipourshiraz et~al\mbox{.}(2024)]%
        {darabipourshiraz2024databites}
\bibfield{author}{\bibinfo{person}{Hasti Darabipourshiraz}, \bibinfo{person}{Dev Ambani}, {and} \bibinfo{person}{Duri Long}.} \bibinfo{year}{2024}\natexlab{}.
\newblock \showarticletitle{DataBites: An embodied and co-creative museum exhibit to foster children's understanding of supervised machine learning}. In \bibinfo{booktitle}{\emph{Proceedings of the 16th Conference on Creativity \& Cognition}}. \bibinfo{pages}{550--555}.
\newblock


\bibitem[Darabipourshiraz et~al\mbox{.}(2026)]%
        {darabipourshiraz2026ai}
\bibfield{author}{\bibinfo{person}{Hasti Darabipourshiraz}, \bibinfo{person}{Lily Murakami~Ng}, \bibinfo{person}{Grace Wang}, \bibinfo{person}{Sophie Rollins}, {and} \bibinfo{person}{Duri Long}.} \bibinfo{year}{2026}\natexlab{}.
\newblock \showarticletitle{AI Unplugged: Exploring Pathways from Physical Simulation to Conceptualization of AI Reasoning Processes}.
\newblock \bibinfo{journal}{\emph{ACM Transactions on Computing Education}} (\bibinfo{year}{2026}).
\newblock


\bibitem[DeVrio et~al\mbox{.}(2025)]%
        {devrio2025taxonomy}
\bibfield{author}{\bibinfo{person}{Alicia DeVrio}, \bibinfo{person}{Myra Cheng}, \bibinfo{person}{Lisa Egede}, \bibinfo{person}{Alexandra Olteanu}, {and} \bibinfo{person}{Su~Lin Blodgett}.} \bibinfo{year}{2025}\natexlab{}.
\newblock \showarticletitle{A Taxonomy of Linguistic Expressions That Contribute To Anthropomorphism of Language Technologies}. In \bibinfo{booktitle}{\emph{Proceedings of the 2025 CHI Conference on Human Factors in Computing Systems}}. \bibinfo{pages}{1--18}.
\newblock


\bibitem[Du et~al\mbox{.}(2024)]%
        {du2024fostering}
\bibfield{author}{\bibinfo{person}{Xiaoxue Du}, \bibinfo{person}{Zhichun Liu}, \bibinfo{person}{Xi Wang}, {and} \bibinfo{person}{Qiping Tang}.} \bibinfo{year}{2024}\natexlab{}.
\newblock \showarticletitle{Fostering AI Literacy through Interactive Game-Based Learning: A Case Study on Enhancing Algorithmic Thinking Skills}. In \bibinfo{booktitle}{\emph{Society for Information Technology \& Teacher Education International Conference}}. Association for the Advancement of Computing in Education (AACE), \bibinfo{pages}{333--338}.
\newblock


\bibitem[Fard and Hawamdeh(2026)]%
        {fard2026growing}
\bibfield{author}{\bibinfo{person}{Fariba Fard} {and} \bibinfo{person}{Suliman Hawamdeh}.} \bibinfo{year}{2026}\natexlab{}.
\newblock \showarticletitle{The Growing Need for AI Literacy in Business and Daily Life}.
\newblock In \bibinfo{booktitle}{\emph{AI-Driven Revolution: Transforming the Business Landscape}}. \bibinfo{publisher}{World Scientific}, \bibinfo{pages}{21--36}.
\newblock


\bibitem[Farina et~al\mbox{.}(2025)]%
        {farina2025ethical}
\bibfield{author}{\bibinfo{person}{Mirko Farina}, \bibinfo{person}{Xiao Yu}, {and} \bibinfo{person}{Andrea Lavazza}.} \bibinfo{year}{2025}\natexlab{}.
\newblock \showarticletitle{Ethical Considerations and Policy Interventions Concerning the Impact of Generative AI Tools in the Economy and in Society}.
\newblock \bibinfo{journal}{\emph{AI and Ethics}} \bibinfo{volume}{5}, \bibinfo{number}{1} (\bibinfo{year}{2025}), \bibinfo{pages}{737--745}.
\newblock


\bibitem[Gebre(2022)]%
        {gebre2022conceptions}
\bibfield{author}{\bibinfo{person}{Engida Gebre}.} \bibinfo{year}{2022}\natexlab{}.
\newblock \showarticletitle{Conceptions and perspectives of data literacy in secondary education}.
\newblock \bibinfo{journal}{\emph{British Journal of Educational Technology}} \bibinfo{volume}{53}, \bibinfo{number}{5} (\bibinfo{year}{2022}), \bibinfo{pages}{1080--1095}.
\newblock


\bibitem[Giannakos et~al\mbox{.}(2020)]%
        {giannakos_games_2020}
\bibfield{author}{\bibinfo{person}{Michail Giannakos}, \bibinfo{person}{Iro Voulgari}, \bibinfo{person}{Sofia Papavlasopoulou}, \bibinfo{person}{Zacharoula Papamitsiou}, {and} \bibinfo{person}{Georgios Yannakakis}.} \bibinfo{year}{2020}\natexlab{}.
\newblock \showarticletitle{Games for {Artificial} {Intelligence} and {Machine} {Learning} {Education}: {Review} and {Perspectives}}.
\newblock In \bibinfo{booktitle}{\emph{Non-{Formal} and {Informal} {Science} {Learning} in the {ICT} {Era}}}, \bibfield{editor}{\bibinfo{person}{Michail Giannakos}} (Ed.). \bibinfo{publisher}{Springer Singapore}, \bibinfo{address}{Singapore}, \bibinfo{pages}{117--133}.
\newblock
\showISBNx{9789811567469 9789811567476}
\href{https://doi.org/10.1007/978-981-15-6747-6_7}{doi:\nolinkurl{10.1007/978-981-15-6747-6_7}}
\newblock
\shownote{Series Title: Lecture Notes in Educational Technology}.


\bibitem[Laupichler et~al\mbox{.}(2022)]%
        {laupichler2022artificial}
\bibfield{author}{\bibinfo{person}{Matthias~Carl Laupichler}, \bibinfo{person}{Alexandra Aster}, \bibinfo{person}{Jana Schirch}, {and} \bibinfo{person}{Tobias Raupach}.} \bibinfo{year}{2022}\natexlab{}.
\newblock \showarticletitle{Artificial intelligence literacy in higher and adult education: A scoping literature review}.
\newblock \bibinfo{journal}{\emph{Computers and Education: Artificial Intelligence}}  \bibinfo{volume}{3} (\bibinfo{year}{2022}), \bibinfo{pages}{100101}.
\newblock


\bibitem[Li et~al\mbox{.}(2025)]%
        {li2025surprising}
\bibfield{author}{\bibinfo{person}{Jiaxuan Li}, \bibinfo{person}{Mingxing Han}, {and} \bibinfo{person}{Qinjian Yuan}.} \bibinfo{year}{2025}\natexlab{}.
\newblock \showarticletitle{The Surprising Paradox of AI Literacy: How Lower AI Literacy Can Lead to Higher Acceptance}.
\newblock \bibinfo{journal}{\emph{International Journal of Human--Computer Interaction}} (\bibinfo{year}{2025}), \bibinfo{pages}{1--21}.
\newblock


\bibitem[Liang et~al\mbox{.}(2025)]%
        {liang2025quantifying}
\bibfield{author}{\bibinfo{person}{Weixin Liang}, \bibinfo{person}{Yaohui Zhang}, \bibinfo{person}{Zhengxuan Wu}, \bibinfo{person}{Haley Lepp}, \bibinfo{person}{Wenlong Ji}, \bibinfo{person}{Xuandong Zhao}, \bibinfo{person}{Hancheng Cao}, \bibinfo{person}{Sheng Liu}, \bibinfo{person}{Siyu He}, \bibinfo{person}{Zhi Huang}, {et~al\mbox{.}}} \bibinfo{year}{2025}\natexlab{}.
\newblock \showarticletitle{Quantifying large language model usage in scientific papers}.
\newblock \bibinfo{journal}{\emph{Nature Human Behaviour}} (\bibinfo{year}{2025}), \bibinfo{pages}{1--11}.
\newblock


\bibitem[Lintner(2024)]%
        {lintner2024systematic}
\bibfield{author}{\bibinfo{person}{Tom{\'a}{\v{s}} Lintner}.} \bibinfo{year}{2024}\natexlab{}.
\newblock \showarticletitle{A systematic review of AI literacy scales}.
\newblock \bibinfo{journal}{\emph{npj Science of Learning}} \bibinfo{volume}{9}, \bibinfo{number}{1} (\bibinfo{year}{2024}), \bibinfo{pages}{50}.
\newblock


\bibitem[Long et~al\mbox{.}(2021)]%
        {long2021co}
\bibfield{author}{\bibinfo{person}{Duri Long}, \bibinfo{person}{Takeria Blunt}, {and} \bibinfo{person}{Brian Magerko}.} \bibinfo{year}{2021}\natexlab{}.
\newblock \showarticletitle{Co-designing AI literacy exhibits for informal learning spaces}.
\newblock \bibinfo{journal}{\emph{Proceedings of the ACM on Human-Computer Interaction}} \bibinfo{volume}{5}, \bibinfo{number}{CSCW2} (\bibinfo{year}{2021}), \bibinfo{pages}{1--35}.
\newblock


\bibitem[Long and Magerko(2020)]%
        {long2020ai}
\bibfield{author}{\bibinfo{person}{Duri Long} {and} \bibinfo{person}{Brian Magerko}.} \bibinfo{year}{2020}\natexlab{}.
\newblock \showarticletitle{What is AI literacy? Competencies and design considerations}. In \bibinfo{booktitle}{\emph{Proceedings of the 2020 CHI conference on human factors in computing systems}}. \bibinfo{pages}{1--16}.
\newblock


\bibitem[Long et~al\mbox{.}(2022)]%
        {long2022family}
\bibfield{author}{\bibinfo{person}{Duri Long}, \bibinfo{person}{Anthony Teachey}, {and} \bibinfo{person}{Brian Magerko}.} \bibinfo{year}{2022}\natexlab{}.
\newblock \showarticletitle{Family learning talk in AI literacy learning activities}. In \bibinfo{booktitle}{\emph{Proceedings of the 2022 CHI Conference on Human Factors in Computing Systems}}. \bibinfo{pages}{1--20}.
\newblock


\bibitem[Ma et~al\mbox{.}(2025)]%
        {ma2025imaginaition}
\bibfield{author}{\bibinfo{person}{Qianou Ma}, \bibinfo{person}{Anika Jain}, \bibinfo{person}{Jini Kim}, \bibinfo{person}{Megan Chai}, {and} \bibinfo{person}{Geoff Kaufman}.} \bibinfo{year}{2025}\natexlab{}.
\newblock \showarticletitle{ImaginAItion: Promoting Generative AI Literacy Through Game-Based Learning}. In \bibinfo{booktitle}{\emph{Proceedings of the Extended Abstracts of the CHI Conference on Human Factors in Computing Systems}}. \bibinfo{pages}{1--9}.
\newblock


\bibitem[McDaniel et~al\mbox{.}(2025)]%
        {mcdaniel2025artificial}
\bibfield{author}{\bibinfo{person}{Brandon~T McDaniel}, \bibinfo{person}{Fayika~F Nova}, {and} \bibinfo{person}{Jessica~A Pater}.} \bibinfo{year}{2025}\natexlab{}.
\newblock \showarticletitle{Artificial Intelligence in Everyday Family Life: Issues, Applications, and Implications}.
\newblock \bibinfo{journal}{\emph{Family Relations}} \bibinfo{volume}{74}, \bibinfo{number}{3} (\bibinfo{year}{2025}), \bibinfo{pages}{1049--1055}.
\newblock


\bibitem[Nand et~al\mbox{.}(2019)]%
        {nand_engaging_2019}
\bibfield{author}{\bibinfo{person}{Kalpana Nand}, \bibinfo{person}{Nilufar Baghaei}, \bibinfo{person}{John Casey}, \bibinfo{person}{Bashar Barmada}, \bibinfo{person}{Farhad Mehdipour}, {and} \bibinfo{person}{Hai-Ning Liang}.} \bibinfo{year}{2019}\natexlab{}.
\newblock \showarticletitle{Engaging children with educational content via {Gamification}}.
\newblock \bibinfo{journal}{\emph{Smart Learning Environments}} \bibinfo{volume}{6}, \bibinfo{number}{1} (\bibinfo{date}{July} \bibinfo{year}{2019}), \bibinfo{pages}{6}.
\newblock
\showISSN{2196-7091}
\href{https://doi.org/10.1186/s40561-019-0085-2}{doi:\nolinkurl{10.1186/s40561-019-0085-2}}


\bibitem[Ng et~al\mbox{.}(2021)]%
        {ng2021conceptualizing}
\bibfield{author}{\bibinfo{person}{Davy Tsz~Kit Ng}, \bibinfo{person}{Jac Ka~Lok Leung}, \bibinfo{person}{Samuel Kai~Wah Chu}, {and} \bibinfo{person}{Maggie~Shen Qiao}.} \bibinfo{year}{2021}\natexlab{}.
\newblock \showarticletitle{Conceptualizing AI literacy: An exploratory review}.
\newblock \bibinfo{journal}{\emph{Computers and Education: Artificial Intelligence}}  \bibinfo{volume}{2} (\bibinfo{year}{2021}), \bibinfo{pages}{100041}.
\newblock


\bibitem[Ng et~al\mbox{.}(2024)]%
        {ng2024fostering}
\bibfield{author}{\bibinfo{person}{Davy Tsz~Kit Ng}, \bibinfo{person}{Chen Xinyu}, \bibinfo{person}{Jac Ka~Lok Leung}, {and} \bibinfo{person}{Samuel Kai~Wah Chu}.} \bibinfo{year}{2024}\natexlab{}.
\newblock \showarticletitle{Fostering students' AI literacy development through educational games: AI knowledge, affective and cognitive engagement}.
\newblock \bibinfo{journal}{\emph{Journal of computer assisted learning}} \bibinfo{volume}{40}, \bibinfo{number}{5} (\bibinfo{year}{2024}), \bibinfo{pages}{2049--2064}.
\newblock


\bibitem[Plass et~al\mbox{.}(2015)]%
        {plass2015foundations}
\bibfield{author}{\bibinfo{person}{Jan~L Plass}, \bibinfo{person}{Bruce~D Homer}, {and} \bibinfo{person}{Charles~K Kinzer}.} \bibinfo{year}{2015}\natexlab{}.
\newblock \showarticletitle{Foundations of game-based learning}.
\newblock \bibinfo{journal}{\emph{Educational psychologist}} \bibinfo{volume}{50}, \bibinfo{number}{4} (\bibinfo{year}{2015}), \bibinfo{pages}{258--283}.
\newblock


\bibitem[Ridsdale et~al\mbox{.}(2015)]%
        {ridsdale2015strategies}
\bibfield{author}{\bibinfo{person}{Chantel Ridsdale}, \bibinfo{person}{James Rothwell}, \bibinfo{person}{Mike Smit}, \bibinfo{person}{Hossam Ali-Hassan}, \bibinfo{person}{Michael Bliemel}, \bibinfo{person}{Dean Irvine}, \bibinfo{person}{Daniel Kelley}, \bibinfo{person}{Stan Matwin}, {and} \bibinfo{person}{Brad Wuetherick}.} \bibinfo{year}{2015}\natexlab{}.
\newblock \showarticletitle{Strategies and best practices for data literacy education}.
\newblock \bibinfo{journal}{\emph{Knowledge synthesis report}}  \bibinfo{volume}{151} (\bibinfo{year}{2015}), \bibinfo{pages}{10--17}.
\newblock


\bibitem[Rollins et~al\mbox{.}(2024)]%
        {rollins2024knowledge}
\bibfield{author}{\bibinfo{person}{Sophie Rollins}, \bibinfo{person}{Katherine Hancock}, \bibinfo{person}{Jasmin Ali-Diaz}, \bibinfo{person}{Nyssa Shahdadpuri}, {and} \bibinfo{person}{Duri Long}.} \bibinfo{year}{2024}\natexlab{}.
\newblock \showarticletitle{Knowledge Net: Fostering Children’s Understanding of Knowledge Representations Through Creative Making and Embodied Interaction in a Museum Exhibit}. In \bibinfo{booktitle}{\emph{Proceedings of the 16th Conference on Creativity \& Cognition}}. \bibinfo{pages}{470--475}.
\newblock


\bibitem[Rondon et~al\mbox{.}(2013)]%
        {rondon_computer_2013}
\bibfield{author}{\bibinfo{person}{Silmara Rondon}, \bibinfo{person}{Fernanda~Chiarion Sassi}, {and} \bibinfo{person}{Claudia~Regina Furquim De~Andrade}.} \bibinfo{year}{2013}\natexlab{}.
\newblock \showarticletitle{Computer game-based and traditional learning method: a comparison regarding students’ knowledge retention}.
\newblock \bibinfo{journal}{\emph{BMC Medical Education}} \bibinfo{volume}{13}, \bibinfo{number}{1} (\bibinfo{date}{Dec.} \bibinfo{year}{2013}), \bibinfo{pages}{30}.
\newblock
\showISSN{1472-6920}
\href{https://doi.org/10.1186/1472-6920-13-30}{doi:\nolinkurl{10.1186/1472-6920-13-30}}


\bibitem[Sager et~al\mbox{.}(2026)]%
        {sager2026data}
\bibfield{author}{\bibinfo{person}{Marc~T Sager}, \bibinfo{person}{Sarah Miller}, {and} \bibinfo{person}{Zarek Drozda}.} \bibinfo{year}{2026}\natexlab{}.
\newblock \showarticletitle{Data Science Education in US Informal Learning Environments: A Review of the Literature}.
\newblock \bibinfo{journal}{\emph{Journal of Statistics and Data Science Education}} \bibinfo{number}{just-accepted} (\bibinfo{year}{2026}), \bibinfo{pages}{1--28}.
\newblock


\bibitem[Salles et~al\mbox{.}(2020)]%
        {salles2020anthropomorphism}
\bibfield{author}{\bibinfo{person}{Arleen Salles}, \bibinfo{person}{Kathinka Evers}, {and} \bibinfo{person}{Michele Farisco}.} \bibinfo{year}{2020}\natexlab{}.
\newblock \showarticletitle{Anthropomorphism in AI}.
\newblock \bibinfo{journal}{\emph{AJOB neuroscience}} \bibinfo{volume}{11}, \bibinfo{number}{2} (\bibinfo{year}{2020}), \bibinfo{pages}{88--95}.
\newblock


\bibitem[Smilkov et~al\mbox{.}(2017)]%
        {smilkov2017direct}
\bibfield{author}{\bibinfo{person}{Daniel Smilkov}, \bibinfo{person}{Shan Carter}, \bibinfo{person}{D Sculley}, \bibinfo{person}{Fernanda~B Vi{\'e}gas}, {and} \bibinfo{person}{Martin Wattenberg}.} \bibinfo{year}{2017}\natexlab{}.
\newblock \showarticletitle{Direct-manipulation visualization of deep networks}.
\newblock \bibinfo{journal}{\emph{arXiv preprint arXiv:1708.03788}} (\bibinfo{year}{2017}).
\newblock


\bibitem[Su et~al\mbox{.}(2023)]%
        {su2023artificial}
\bibfield{author}{\bibinfo{person}{Jiahong Su}, \bibinfo{person}{Davy Tsz~Kit Ng}, {and} \bibinfo{person}{Samuel Kai~Wah Chu}.} \bibinfo{year}{2023}\natexlab{}.
\newblock \showarticletitle{Artificial intelligence (AI) literacy in early childhood education: The challenges and opportunities}.
\newblock \bibinfo{journal}{\emph{Computers and Education: Artificial Intelligence}}  \bibinfo{volume}{4} (\bibinfo{year}{2023}), \bibinfo{pages}{100124}.
\newblock


\bibitem[Tobias et~al\mbox{.}(2013)]%
        {tobias2013game}
\bibfield{author}{\bibinfo{person}{Sigmund Tobias}, \bibinfo{person}{J~Dexter Fletcher}, {and} \bibinfo{person}{Alexander~P Wind}.} \bibinfo{year}{2013}\natexlab{}.
\newblock \showarticletitle{Game-based learning}.
\newblock \bibinfo{journal}{\emph{Handbook of research on educational communications and technology}} (\bibinfo{year}{2013}), \bibinfo{pages}{485--503}.
\newblock


\bibitem[Trujillo et~al\mbox{.}(2025)]%
        {trujillo2025global}
\bibfield{author}{\bibinfo{person}{Herminio~Pab{\'o}n Trujillo}, \bibinfo{person}{Juan~Carlos Castillo}, \bibinfo{person}{Angie~Dayana Rangel}, \bibinfo{person}{Oscar~Fabi{\'a}n Pati{\~n}o}, \bibinfo{person}{Ronald Edinxon~Angarita Bautista}, {and} \bibinfo{person}{Jenny Zulay~C{\'a}ceres C{\'a}rdenas}.} \bibinfo{year}{2025}\natexlab{}.
\newblock \showarticletitle{The Global Impact of AI on Workplace Safety, Opportunities and Challenges for the Future of Work}.
\newblock \bibinfo{journal}{\emph{Ciencia Latina Revista Cient{\'\i}fica Multidisciplinar}} \bibinfo{volume}{9}, \bibinfo{number}{2} (\bibinfo{year}{2025}), \bibinfo{pages}{7500--7513}.
\newblock


\bibitem[Tully et~al\mbox{.}(2025)]%
        {tully2025lower}
\bibfield{author}{\bibinfo{person}{Stephanie~M Tully}, \bibinfo{person}{Chiara Longoni}, {and} \bibinfo{person}{Gil Appel}.} \bibinfo{year}{2025}\natexlab{}.
\newblock \showarticletitle{Lower artificial intelligence literacy predicts greater AI receptivity}.
\newblock \bibinfo{journal}{\emph{Journal of Marketing}} (\bibinfo{year}{2025}), \bibinfo{pages}{00222429251314491}.
\newblock


\bibitem[Wang et~al\mbox{.}(2025)]%
        {wang2025does}
\bibfield{author}{\bibinfo{person}{Qiang Wang}, \bibinfo{person}{Tingting Sun}, {and} \bibinfo{person}{Rongrong Li}.} \bibinfo{year}{2025}\natexlab{}.
\newblock \showarticletitle{Does Artificial Intelligence (AI) Enhance Green Economy Efficiency? The Role of Green Finance, Trade Openness, and R\&D Investment}.
\newblock \bibinfo{journal}{\emph{Humanities and Social Sciences Communications}} \bibinfo{volume}{12}, \bibinfo{number}{1} (\bibinfo{year}{2025}), \bibinfo{pages}{1--22}.
\newblock


\bibitem[Wood et~al\mbox{.}(2025)]%
        {wood2025exploratory}
\bibfield{author}{\bibinfo{person}{Geert Wood}, \bibinfo{person}{Elena Nu{\~n}ez~Castellar}, {and} \bibinfo{person}{Wijnand IJsselsteijn}.} \bibinfo{year}{2025}\natexlab{}.
\newblock \showarticletitle{An Exploratory Study Into the Impact of AI Literacy Training on Anthropomorphism and Trust in Conversational AI}. In \bibinfo{booktitle}{\emph{International Conference on Human-Computer Interaction}}. Springer, \bibinfo{pages}{301--322}.
\newblock


\bibitem[Zeng et~al\mbox{.}(2020)]%
        {zeng_learn_2020}
\bibfield{author}{\bibinfo{person}{Jialing Zeng}, \bibinfo{person}{Sophie Parks}, {and} \bibinfo{person}{Junjie Shang}.} \bibinfo{year}{2020}\natexlab{}.
\newblock \showarticletitle{To learn scientifically, effectively, and enjoyably: {A} review of educational games}.
\newblock \bibinfo{journal}{\emph{Human Behavior and Emerging Technologies}} \bibinfo{volume}{2}, \bibinfo{number}{2} (\bibinfo{year}{2020}), \bibinfo{pages}{186--195}.
\newblock
\showISSN{2578-1863}
\href{https://doi.org/10.1002/hbe2.188}{doi:\nolinkurl{10.1002/hbe2.188}}


\end{thebibliography}

\end{document}